\documentclass[aps,prl,twocolumn,superscriptaddress,floatfix]{revtex4}
\usepackage{graphicx}
\usepackage{amssymb}

\begin{document}

\title{Effect of Fermi Surface Nesting on Resonant Spin Excitations in Ba$_{1-x}$K$_{x}$Fe$_2$As$_2$}

\author{J.-P. Castellan}
\author{S. Rosenkranz}
\author{E. A. Goremychkin}
\author{D. Y. Chung}
\author{I. S. Todorov}
\affiliation{Materials Science Division, Argonne National Laboratory, Argonne, IL 60439-4845, USA}
\author{M. G. Kanatzidis}
\affiliation{Materials Science Division, Argonne National Laboratory, Argonne, IL 60439-4845, USA}
\affiliation{Department of Chemistry, Northwestern University, Evanston, IL 60208-3113, USA}
\author{I. Eremin}
\affiliation{Institute for Theoretical Physics III, Ruhr University Bochum, 44801 Bochum, Germany}
\author{J. Knolle}
\affiliation{Max-Planck-Institut f\"{u}r Physik komplexer Systeme, D-01187 Dresden, Germany}
\author{A. V. Chubukov}
\author{S. Maiti}
\affiliation{Department of Physics, University of Wisconsin-Madison, Madison,Wisconsin 53706, USA}
\author{M. R. Norman}
\author{F. Weber\footnote{Current Address: Karlsruhe Institute of Technology, Institute of Solid State Physics, 76021 Karlsruhe, Germany}}
\author{H. Claus}
\affiliation{Materials Science Division, Argonne National Laboratory, Argonne, IL 60439-4845, USA}
\author{T. Guidi}
\author{R. I. Bewley}
\affiliation{ISIS Pulsed Neutron and Muon Facility, Rutherford Appleton Laboratory, Chilton, Didcot OX11 0QX, United Kingdom}
\author{R. Osborn}
\affiliation{Materials Science Division, Argonne National Laboratory, Argonne, IL 60439-4845, USA}
\email{ROsborn@anl.gov}

\begin{abstract}
We report inelastic neutron scattering measurements of the resonant spin excitations in Ba$_{1-x}$K$_{x}$Fe$_2$As$_2$ over a broad range of electron band filling. The fall in the superconducting transition temperature with hole doping coincides with the magnetic excitations splitting into two incommensurate peaks because of the growing mismatch in the hole and electron Fermi surface volumes, as confirmed by a tight-binding model with $s_{\pm}$-symmetry pairing.  The reduction in Fermi surface nesting is accompanied by a collapse of the resonance binding energy and its spectral weight caused by the weakening of electron-electron correlations.
\end{abstract}

\date{\today}

\maketitle

The connection between magnetism and unconventional superconductivity is one of the most challenging issues in condensed matter physics. In unconventional superconductors, such as the copper oxides\cite{Bonn:2006p34943}, heavy fermions\cite{Pfleiderer:2009p31671}, organic charge-transfer salts\cite{McKenzie:1997p34926}, and now the iron pnictides and chalcogenides\cite{DeLaCruz:2008p8095,Lynn:2009p28738,Taillefer:2010p34241}, the superconducting state occurs in the presence of strong magnetic correlations and sometimes coexists with magnetic order, fostering models of superconducting pairing mediated by magnetic fluctuations\cite{Mazin:2008p11687,Dahm:2009p15572,Yu:2009p31175}. Such models lead to unusual superconducting gap symmetries, such as the $d$-wave symmetry observed in the high-temperature copper oxide superconductors. Although there have been reports of energy gap nodes in a few of the iron superconductors, most appear to have weakly anisotropic gaps\cite{Paglione:2010p34248}. This is consistent both with conventional $s$-wave pairing, in which the gap has the same sign over the entire Fermi surface, and with unconventional $s_{\pm}$-wave pairing, in which the gaps on the disconnected hole and electron Fermi surfaces have opposite sign\cite{Mazin:2008p11687}.

The first spectroscopic evidence of unconventional $s_{\pm}$-wave symmetry was provided by inelastic neutron scattering on optimally-doped Ba$_{0.6}$K$_{0.4}$Fe$_2$As$_2$ with the observation of a resonant spin excitation at the wavevector, $\mathbf{Q}$, connecting the nearly cylindrical hole and electron Fermi surfaces, centered at the zone center ($\Gamma$-point) and zone boundary (M-point), respectively, \textit{i.e.}, at $\mathbf{Q}_0=(\pi,\pi)$ in the crystallographic Brillouin zone\cite{Christianson:2008p14965}. Similar excitations have now been observed in a wide range of iron-based superconductors \cite{Chi:2009p20185,Lumsden:2009p20184,Inosov:2009p31864,Lumsden:2010p31952}. Within an itinerant model, the resonance arises from an enhancement in the superconducting phase of the band electron susceptibility, $\chi(\mathbf{Q},\omega)$, caused by coherence factors introduced by pair formation\cite{Korshunov:2008p13468,Maier:2009p21834}. If $\Delta_{\mathbf{k}+\mathbf{Q}}=-\Delta_{\mathbf{k}}$, where $\Delta_{\mathbf{k}}$ are the values of the energy gap at points $\mathbf{k}$ on the Fermi surface, \textit{i.e.}, if  \textbf{Q} connects points whose gaps have opposite sign, the magnetic susceptibility of superconducting but otherwise non-interacting fermions Re\,$\chi_0(\mathbf{Q},\omega)$ diverges logarithmically upon approaching $2\Delta$ from below. In the Random Phase Approximation (RPA), the full susceptibility is given by 
\begin{equation}
\chi(\mathbf{Q},\omega)=\chi_0(\mathbf{Q},\omega)\left[1-J(\mathbf{Q})\chi_0(\mathbf{Q},\omega)\right]^{-1}
\end{equation}
where $J(\mathbf{Q})$ represents electron-electron interactions. The interactions produce a bound exciton with an energy $\Omega$ below $2\Delta$ given by $J(\mathbf{Q})\mathrm{Re}\,\chi_0(\mathbf{Q},\Omega) =1$, \textit{i.e.} with a binding energy of $2\Delta - \Omega$\cite{Eschrig:2006p23369}. 

\begin{figure*}[!htb]
\centering
\includegraphics[width=1.8\columnwidth]{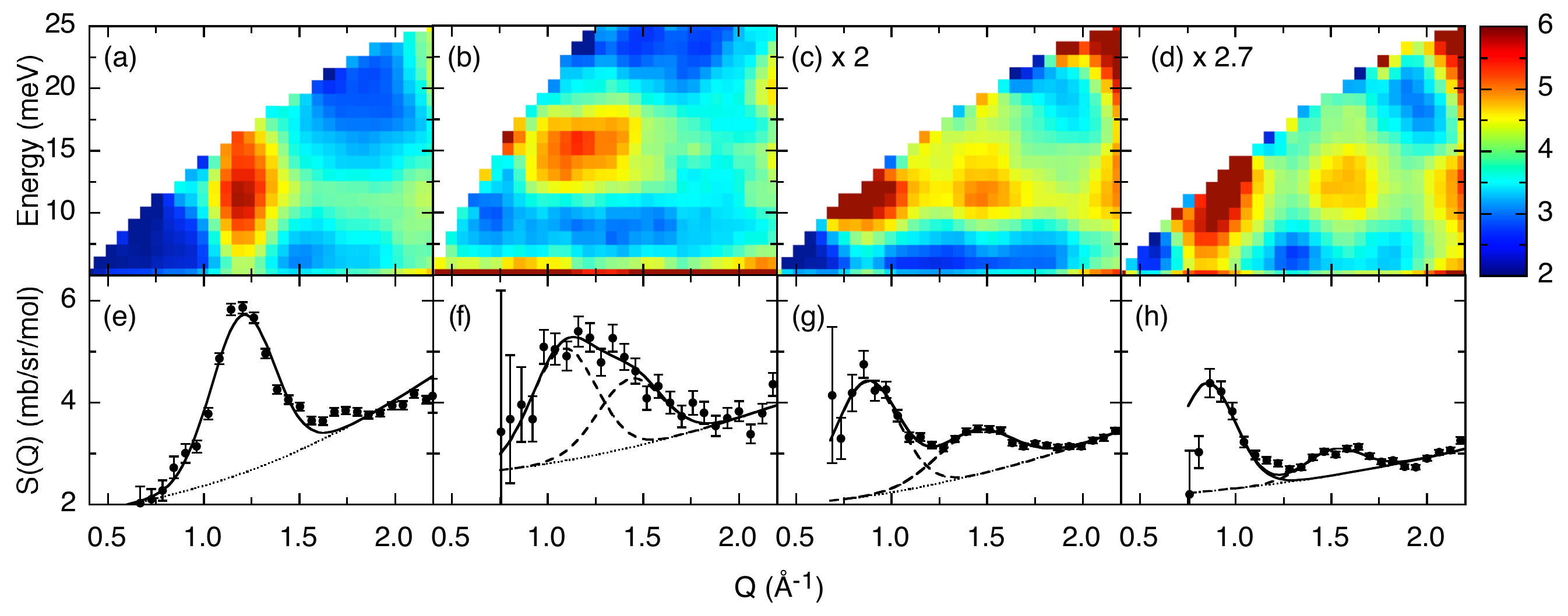}
\vspace{-0.15in}
\caption{(a-d) Inelastic neutron scattering from Ba$_{1-x}$K$_{x}$Fe$_2$As$_2$ measured in the superconducting phase at a temperature of 5\,K using incident neutron energies (E$_i$) of 30\,meV and 60\,meV, at (a) $x=0.3$ (E$_i$=30\,meV), (b) $x=0.5$ (E$_i$=60\,meV), (c) $x=0.7$ (E$_i$=30\,meV), scaled by a factor 2.0, and (d) $x=0.9$ (E$_i$=30\,meV), scaled by a factor 2.667.  (e-h) The magnetic scattering \textit{vs Q} integrated over the energy transfer range of the inelastic peak, compared to a fitted model that includes one (e) or two Gaussian peaks (f-h) (dashed black lines) and a non-magnetic background (dotted line), given by the sum of a quadratic term, consistent with single-phonon scattering, and a constant term, consistent with multiple phonon scattering. The energy integration range is (e) 9-14\,meV, (f) 12-18\,meV, (g) 10-14\,meV and (h) 10-15\,meV.
\label{Fig1} }
\vspace{-0.2in}
\end{figure*}

This letter addresses the evolution of the resonant spin excitations with band filling. Resonant spin excitations in the iron-based superconductors have mostly been observed in compounds close to optimal doping where the hole and electron pockets have similar size. We have now studied the magnetic excitations in Ba$_{1-x}$K$_{x}$Fe$_2$As$_2$ over a broad range of hole dopings in which the mismatch between the hole and electron Fermi surface volumes becomes increasingly significant. Our results therefore provide insight into the influence of Fermi surface nesting on the unconventional superconductivity. At moderate doping, there is a longitudinal broadening of the wavevector of the magnetic response and then, at higher doping, a split into two incommensurate peaks, which is correlated with a fall in T$_c$. The scaling of the resonant peak energy to the maximum energy of the superconducting gap, $\Omega/2\Delta$, is not universal as has been claimed\cite{Yu:2009p31175}, but renormalizes continuously to 1 with increasing hole concentration. This represents a reduction in the exciton binding energy due to weakening electron-electron interactions and is accompanied by a collapse of the resonant spectral weight\cite{Eschrig:2006p23369}.

The reason for choosing Ba$_{1-x}$K$_{x}$Fe$_2$As$_2$ in our investigation is that the superconducting phase extends over a much broader range of dopant concentration with hole-doping $(0.125\leq x\leq 1.0)$ than with electron-doping $(0.08\leq x\leq 0.32)$\cite{Canfield:2010p34239}. According to ARPES data\cite{Sato:2009p24879}, the shift of the chemical potential from optimal doping at $x=0.4$ to the extreme overdoping at $x=1$ approximately doubles the radius of the hole pockets and shrinks the electron pockets, which vanish close to $x=1$. 

We have prepared polycrystalline samples with $x=0.3$, 0.5, 0.7, and 0.9, to supplement our earlier measurements of $x=0.4$. Details of the sample synthesis procedures are reported elsewhere\cite{Avci:2011p36647}.  The inelastic neutron measurements were performed on the Merlin spectrometer at the ISIS Pulsed Neutron Facility, UK, using incident energies of 30 and 60\,meV and temperatures of 5\,K and 50\,K, \textit{i.e.}, below and above T$_\mathrm{c}$. The data were placed on an absolute intensity scale by normalization to a vanadium standard. 

\begin{figure*}[!htb]
\centering
\includegraphics[width=1.6\columnwidth]{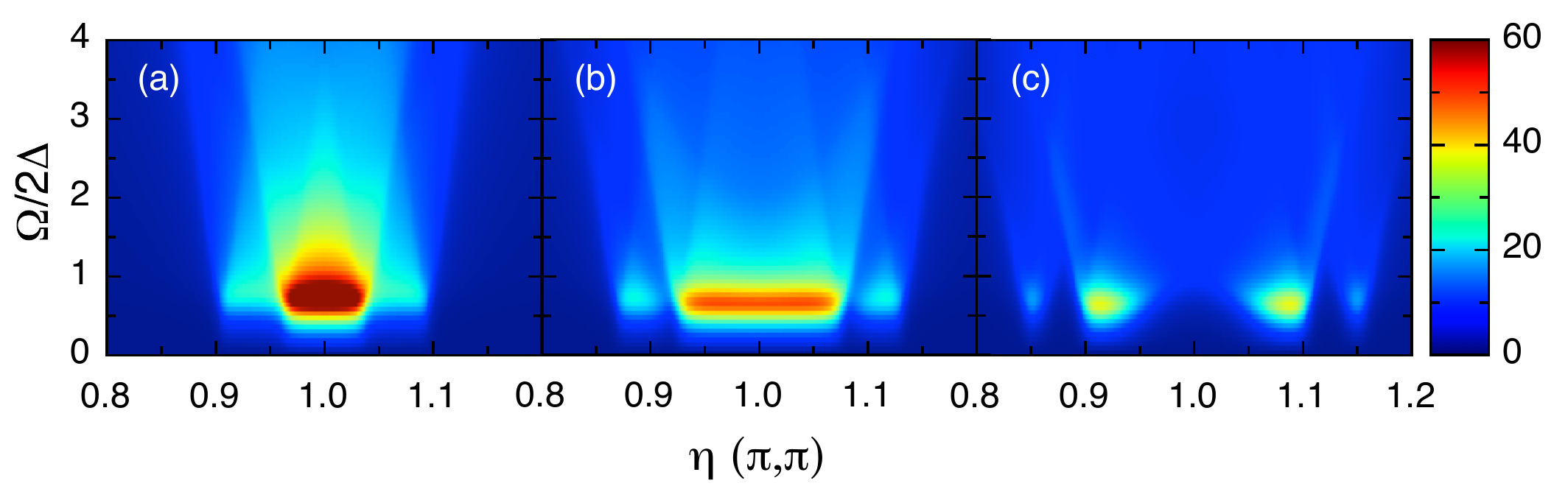}
\vspace{-0.15in}
\caption{Calculated Im$\chi(Q,\omega)$ with increasing hole concentration based on a four-band tight-binding model with two circular hole pockets and two elliptical electron pockets, with ellipticity $\epsilon=0.5$ and chemical potentials (a) $\mu=0.0$, (b) $\mu=0.3$, and (c) $\mu=0.5$. The intensity map corresponds to states/eV.
\label{Fig2} }
\vspace{-0.2in}
\end{figure*}

Fig. 1(a-d) summarizes the data at low temperature showing that inelastic peaks are visible in all compositions centered in energy between $\sim11$\,meV and 15\,meV. The most striking observation is the pronounced $Q$-broadening of the inelastic scattering at $x=0.5$ and its split into two incommensurate peaks at $x=0.7$ and 0.9. This is seen most clearly in Fig. 1(e-h), which shows the wavevector dependence of the energy-integrated intensity. The magnetic contribution is fit to one or two peaks symmetrically centered around $Q_0\sim1.2$\,\AA$^{-1}$, with phonons and multiple scattering contributing a quadratic $Q$-dependent and $Q$-independent intensity, respectively. After correction for the Fe$^{2+}$ form factor, the split peaks have approximately equal intensity. Since we are measuring polycrystalline samples, the $l$-dependence is spherically averaged, but this cannot explain the size of the splitting. The absolute values of $Q$ at $(\pi,\pi,0)$ and $(\pi,\pi,\pm\pi)$ are 1.15\,\AA$^{-1}$ to 1.25\,\AA$^{-1}$, respectively, whereas the peaks at $x=0.7$ are at 0.94\,\AA$^{-1}$ and 1.52\,\AA$^{-1}$. 

In order to understand the incommensurability, we have performed calculations of the doping dependence of the dynamic magnetic susceptibility using a simple tight-binding model of two hole pockets centered around the $\Gamma$-point and two electron pockets centered around the M-points. To make quantitative as well as qualitative comparisons to experiments, we use the Fermi velocities and the size of the Fermi pockets based on Refs. \cite{Singh:2008p8173,Mazin:2008p11687}. A key parameter is the ellipticity, $\epsilon$, of the electron pockets. Perfect nesting requires $\epsilon$ to be zero, \textit{i.e.}, circular electron pockets. With increasing ellipticity, the magnetic peak broadens and extends to larger $\Delta Q$ around $Q_0$, because the increasing mismatch of the hole and electron Fermi surfaces weakens the singularity in the magnetic response at $Q_0$. The intensity is also lower because of the weaker nesting and smaller size of the superconducting gap on this Fermi surface.

In the following, we set $\epsilon=0.5$, which gives the best description of the spin waves in the parent compound\cite{Knolle:2010p33082}, to investigate the doping evolution of Im$\chi(Q,\omega)$ for various hole dopings, \textit{i.e.} for positive values of the hole doping parameter, $\mu$. In agreement with the experiments, the magnetic response is initially commensurate and a well-resolved splitting is only found at $\mu=0.5$. The critical doping at which the magnetic peak splits, $\mu_c$, also depends on the ellipticity. For larger (smaller) $\epsilon$, the splitting becomes visible at larger (smaller) values of $\mu$. 

A comparison of Fig. 1 and Fig. 2 demonstrates that the calculations reproduce the observed behavior as a function of potassium doping. The values of the calculated incommensurability \textit{vs} $\mu$ are plotted with the experimental data in Fig. 3a, assuming $\mu\sim x$. The good agreement, with $\mu_c\sim 0.4$, shows that the observed incommensurability is consistent with the change in Fermi surface geometry with hole doping. Furthermore, the value of the incommensurability at $x=0.9$ is in agreement with neutron results from pure KFe$_2$As$_2$\cite{Lee:2011hm}, which were also interpreted as interband scattering.  Fig. 3b shows that the onset of incommensurability with $x$ occurs just when T$_c$ starts to fall, showing a direct correlation between the degree of Fermi surface nesting and the pairing strength. 

\begin{figure}[!b]
\centering
\includegraphics[width=0.6\columnwidth]{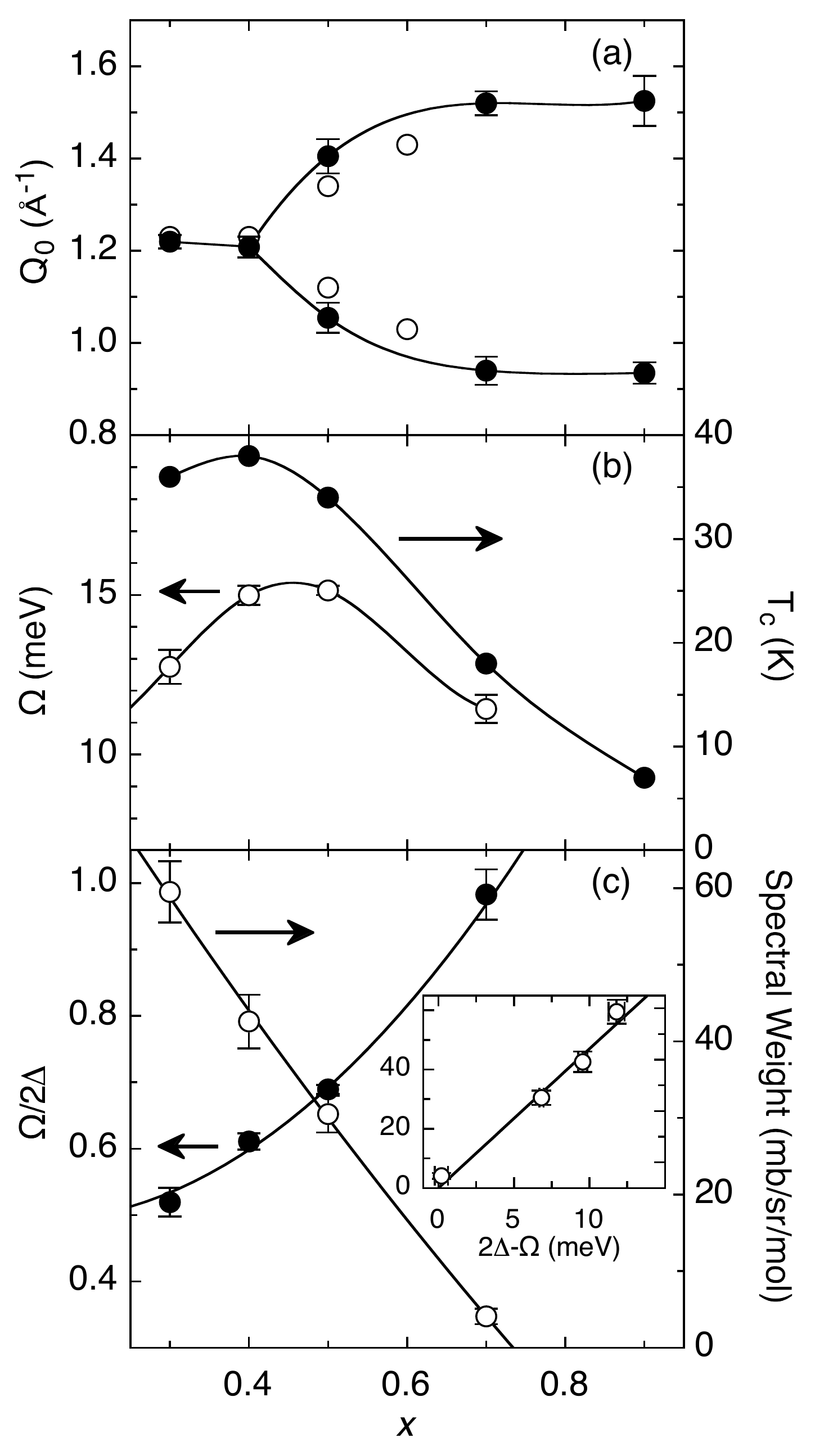}
\vspace{-0.1in}
\caption{(a) The wavevector, $Q$, of the magnetic excitations in Ba$_{1-x}$K$_{x}$Fe$_2$As$_2$, determined from the peak centers in Fig. 1(e-h) \textit{vs} $x$ (solid circles). The open circles are from the theoretical calculations in Fig. 2 assuming $\mu=x$. (b) T$_\mathrm{c}$ (solid circles) and $\Omega$ (open circles) determined from the resonantly enhanced component of the inelastic peaks shown in Fig. 1, \textit{i.e.}, after subtracting the 50\,K data from the 5\,K data. (c) The ratio of the resonant excitation energy to twice the maximum superconducting energy gap, $\Omega/2\Delta$ (solid circles), using $2\Delta/\mathrm{k_B T_c}=7.5$,\cite{Nakayama:2011eh}, and the resonant spectral weight (open circles). The inset shows the linear dependence of the spectral weight \textit{vs} $2\Delta-\Omega$. All other lines are guides to the eye.
\label{Fig3} }
\vspace{-0.1in}
\end{figure}

The doping dependence of the inelastic peak energies, $\Omega$, is summarized in Fig. 3b, where they are plotted \textit{vs} $x$, along with T$_\mathrm{c}$, which falls from 38\,K at $x=0.4$ to 7\,K at $x=0.9$. ARPES measurements  suggest that the gap scales as $2\Delta/\mathrm{k_B T_c}\sim7.5$\cite{Nakayama:2011eh}. With this assumption, $\Omega/2\Delta$ is observed to increase continuously from 0.52 at $x=0.3$ to 0.98 at $x=0.7$ (Fig. 3c). This is clearly inconsistent with the postulated universal scaling of $\Omega/2\Delta\sim0.64$\cite{Yu:2009p31175}, proposed to characterize all unconventional superconductors, including the copper oxides and heavy fermions. 

\begin{figure}[!htb]
\centering
\includegraphics[width=0.75\columnwidth]{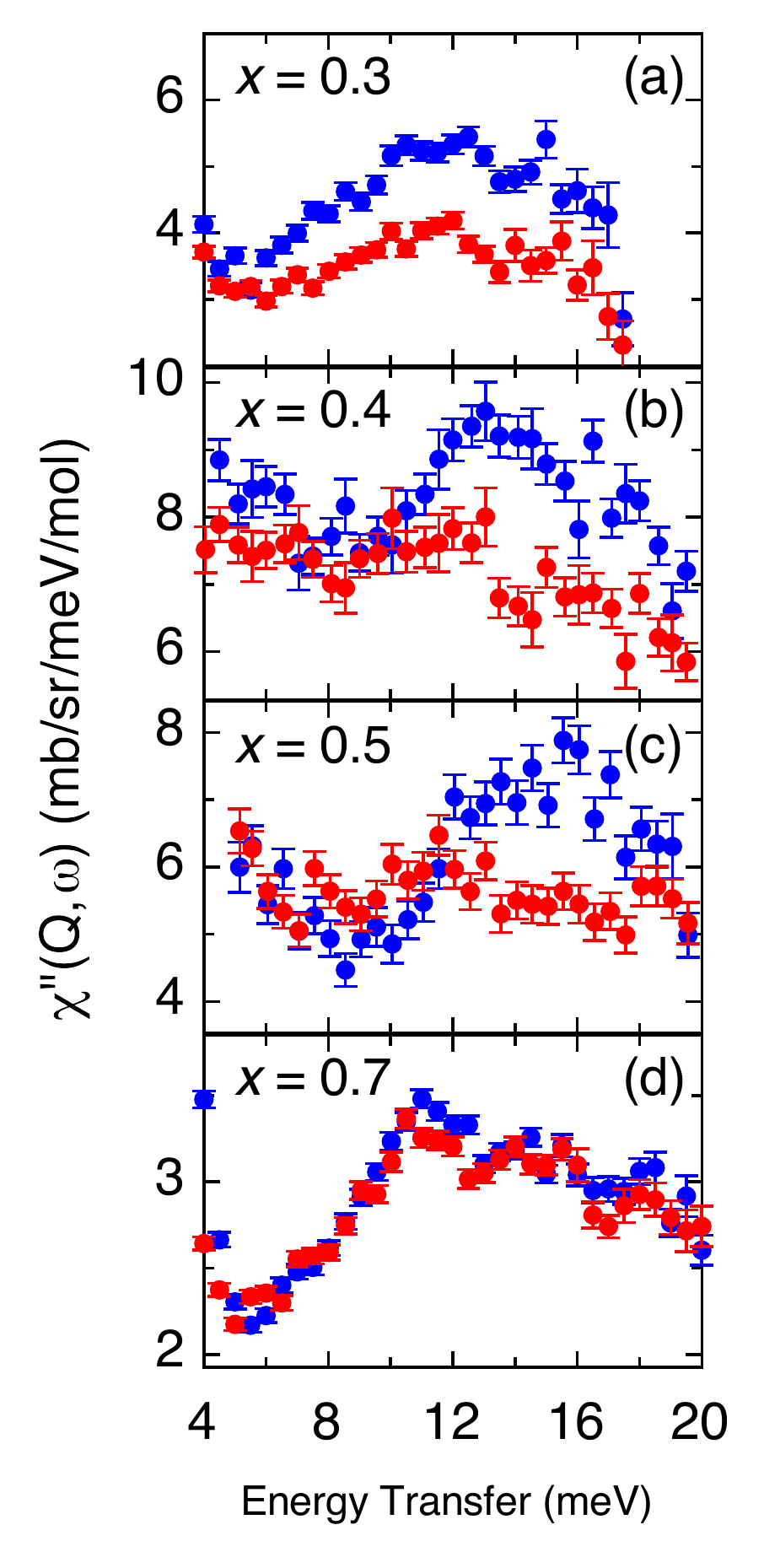}
\vspace{-0.2in}
\caption{Inelastic neutron scattering from Ba$_{1-x}$K$_{x}$Fe$_2$As$_2$ \textit{vs} energy transfer in meV measured at a temperature of 5\,K (blue circles) and 50\,K (red circles) using incident neutron energies (E$_i$) of 30\,meV and 60\,meV. (a) $x=0.3$ (E$_i$=30\,meV), (b) $x=0.4$ (E$_i$=60\,meV) from Ref. \cite{Christianson:2008p14965}, (c) $x=0.5$ (E$_i$=60\,meV), and (d) $x=0.7$ (E$_i$=30\,meV). The $Q$-integration ranges are (a,b,c) 1.0 to 1.4\,\AA$^{-1}$  and (d) 1.2 to 2.0\,\AA$^{-1}$, \textit{i.e.} only the peak at higher-$Q$ is included for $x=0.7$ so the data are plotted on an expanded scale to correct for this and the reduction in Fe$^{2+}$ form factor. The resonant enhancement at $x=0.7$ (d), is also observed in the lower-$Q$ peak.
\label{Fig4} }
\vspace{-0.2in}
\end{figure}

At $x=0.9$, where the magnetic scattering peaks at $\sim13$\,meV, $\Omega/2\Delta$ would be greater than 1, which is inconsistent with the requirement that the resonance is a bound state with a maximum energy of $2\Delta$. In order to explain this anomaly, it is necessary to look at how much of the observed magnetic spectral weight is enhanced in the superconducting phase. Fig. 4 shows the energy spectra for $0.3\leq x \leq 0.7$, including $x=0.4$ published earlier\cite{Christianson:2008p14965}, at 5\,K and 50\,K, \textit{i.e.}, both below and above T$_\mathrm{c}$. The data have been converted to Im$\chi(Q,\omega)$ by correcting for the Bose temperature factor. At $x=0.3$, 0.4, and 0.5, the resonant enhancement of the intensity below T$_\mathrm{c}$ is clearly evident, but at $x=0.7$, it is only just statistically significant.  Fig. 4(d) shows the resonant enhancement in the high-Q peak, but it is also evident in the low-Q peak. Fig. 3(c) shows that the resonant spectral weight determined by subtracting the 50\,K data from the 5\,K data decreases sharply with increasing $x$, falling to zero at $x\sim0.72$. We had insufficient time to measure the $x=0.9$ spectra above T$_\mathrm{c}$, but the trend at lower $x$ suggests that there would be no resonant enhancement below T$_\mathrm{c}$.

This collapse in the resonant spectral weight is clearly linked to the increase of $\Omega/2\Delta$ to 1 (Fig. 3c), a correlation that is predicted by RPA models developed to explain neutron scattering results in the copper oxide superconductors\cite{Eschrig:2006p23369,Pailhes:2006p32258}. In Equation 1, the precise value of the resonance energy is dictated by the interaction term $J$, so $\Omega/2\Delta$ is predicted to shift towards 1 as the electron correlations weaken. This was the interpretation of point-contact tunneling on overdoped Bi$_2$Sr$_2$CaCu$_2$O$_{8+\delta}$\cite{Zasadzinski:2001p32155} and it is quite plausible that electron correlations are diminished in Ba$_{1-x}$K$_{x}$Fe$_2$As$_2$ as Fermi surface nesting and the consequent antiferromagnetic correlations are weakened by hole doping. The itinerant models predict that the reduction in the resonant spectral weight is directly proportional to the reduction in the exciton binding energy, $2\Delta-\Omega$\cite{Abanov:2001p34962}, in excellent agreement with the linear relation shown in the inset to Fig. 3c.

The resonant spin excitations in Ba$_{1-x}$K$_{x}$Fe$_2$As$_2$ prove to be sensitive probes of both the symmetry of the superconducting gap and the Fermi surface geometry.  The incommensurability is seen to be a signature of imperfect nesting, and its onset with increasing $x$ is correlated with the decline in T$_\mathrm{c}$ and the collapse of the resonant spectral weight. The close correspondence between the strength of Fermi surface nesting and superconductivity lends considerable weight to models in which magnetic fluctuations provide the `pairing glue' in the iron pnictide and chalcogenide superconductors.

This work was supported by the Materials Sciences and Engineering Division of the Office of Basic Energy Sciences, Office of Science, U.S. Department of Energy, under contract No. DE-AC02-06CH11357.

%\bibliography{BaKFe2As2doping}

\end{document}